\documentclass[journal]{IEEEtran}

\usepackage{graphicx,amssymb,amsmath,color,textcomp,bm,cite,stfloats}
\usepackage[hidelinks]{hyperref}
\newtheorem{remark}{\underline{Remark}}

\newcounter{MYtempeqncnt}
\begin{document} 

\title{ 
  Doppler  Shift and  Channel 
Estimation for Intelligent Transparent Surface Assisted  Communication 
Systems    on  High-Speed Railways 
\thanks{ Yirun Wang is with the School of Information and Electronics, Beijing
Institute of Technology, 
Beijing 100081, China   
(email: yrwang719@163.com).  } 
\thanks{ Gongpu Wang is with the Beijing Key Laboratory of Transportation Data Analysis and Mining, Beijing Jiaotong University, Beijing 100044,  China   
(e-mail: gpwang@bjtu.edu.cn).  }  
\thanks{ Ruisi He and Bo Ai are with the State Key Laboratory of Rail Traffic Control and Safety, Beijing Jiaotong University, Beijing 100044, China 
(email: ruisi.he@bjtu.edu.cn; boai@bjtu.edu.cn).  }
\thanks{ Chintha Tellambura is with the Department of Electrical and Computer
Engineering, University of Alberta,  Edmonton, AB T6G 2V4, Canada 
(e-mail: chintha@ece.ualberta.ca). }
\author{Yirun Wang, Gongpu Wang, {\it Member, IEEE}, Ruisi He, {\it Senior Member, IEEE}, \\
Bo Ai, {\it  Fellow, IEEE}, and Chintha Tellambura, {\it  Fellow,  IEEE}}
}
\maketitle

\begin{abstract}
The critical distinction between the emerging 
intelligent transparent surface (ITS) and intelligent reflection surface (IRS) is that the incident signals  can penetrate the ITS instead of being reflected,  which  enables the ITS to  combat the severe signal penetration loss for  high-speed railway (HSR) wireless communications.
This paper thus investigates the channel estimation problem   for an  ITS-assisted HSR network where the ITS is embedded into the carriage window.  We first formulate the channels as functions of physical parameters, and thus  transform  the channel estimation    into a parameter recovery problem. Next, we design the first two pilot blocks 
within each frame and develop a serial low-complexity channel estimation algorithm. Specifically, the channel estimates are initially obtained, and each estimate is further  expressed as the sum of its perfectly known value and the estimation error. By leveraging the relationship between channels for the two pilot blocks,  we recover the Doppler shifts from the channel estimates, based on which we can further acquire other channel parameters.   Moreover, the Cram\'er-Rao lower bound (CRLB) for each parameter is derived as a performance benchmark. 
Finally, we provide numerical results to establish the effectiveness of our proposed estimators.   
\end{abstract}

\begin{IEEEkeywords}
Channel estimation, 
Cram\'er-Rao lower bound (CRLB), 
high-speed railway (HSR) communications, 
intelligent transparent surface (ITS),
least square (LS).
\end{IEEEkeywords}

\section{Introduction} 
High-speed railway (HSR) greatly  
facilitates the daily commutes of millions of people with less energy consumption and air pollution than conventional transportation systems \cite{HSR_survery_Guan,HSR_survey/LoS_Ai}. 
One of the crucial parts of HSR operations and economic viability is the onboard availability of wireless communication systems, which not only safeguards the train operations  but also delivers modern communication services and applications to passengers. 

Nonetheless, one critical hindrance to HSR communications is the extreme penetration loss  \cite{HSR_challenge}. 
In the train industries, the body of carriages is made of aluminium in general, through which the  electromagnetic (EM) waves suffer from considerable power loss, and thus the carriage windows are the main paths for transmission \cite{HSR_peloss_Jamaly_Q}. 
Unfortunately, the coated windowpanes are adopted for the purpose of thermal insulation and energy saving \cite{HSR_peloss_Kiani}, 
which result in a sigificant penetration loss of  more than 20  dB to EM waves propagating through windows at the same time \cite{HSR_peloss_Jamaly}. 
Traditional solutions to this issue include relay \cite{HSR_peloss_relay} and multiple-input multiple-output (MIMO) technology \cite{HSR_peloss_MIMO}, both of which however incur high energy consumption, implementation complexity and hardware cost. 
With the rapid development of technology, HSR communication has been evolving into an intelligent era \cite{HSR_smart}, 
where we seek new and promising methods to address the forgoing challenge.

Recently, the intelligent reflection surface (IRS) and its various equivalents have drawn extensive attention as a spectrum- and  energy-efficient technology to smartly reconfigure the radio propagation environment for B5G/6G wireless communication systems 
\cite{IRS_magazine_Wu,IRS_access_Basar}. 
An IRS is generally a planar metasurface comprising numerous passive reflection elements, with each one being programmable to change the amplitude and phase of the incident signals 
\cite{IRS_jointbeam_a/p_Wu,IRS_tutorial_Wu}. 
Therefore, with the help of the IRS, the desirable non-reflected signals can be enhanced by  constructively superimposing them on those reflected ones, or the undesirable signals (e.g.,  electromagnetic interference) can be weakened through destructive combining.  
In addition, IRS also has massive advantages from the implementation aspect, e.g., scalable cost,  lightweight, conformal geometry, excellent compatibility, and others
\cite{IRS_jointbeam_discrete_Wu,IRS_jointbeam_Li/Tao,IRS_journal_Renzo}.

\begin{figure}[t]
\centering
\includegraphics[width=0.8\linewidth]{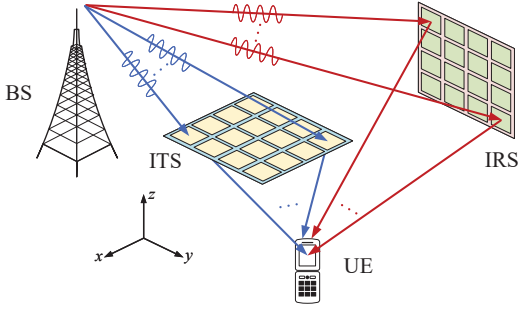}
\caption{Comparison between the refractive ITS and reflective IRS.}
\label{fig:ITSvsIRS}
\end{figure}

The existing works on the IRS-assisted  high-mobility communications mainly consider that the stationary IRS serves the mobile users   
\cite{IRS_mobility_OFDM,IRS_mobility_arXiv,IRS_mobility_network,IRS_mobility_network_VTC,IRS_mobility_network_WCNC}.    
Nevertheless, the signal convergence of passive IRS is much smaller than that of the active transceiver; hence, it is impractical to satisfy the service demands of users during their entire movements.  
To this end, instead of utilizing the reflection capability of the metasurface, we further consider a new type of intelligent manufactured surface comprising multiple passive refraction elements,  named intelligent transparent surface (ITS). 
That is, ITS enables transparent penetration for incident signals while inheriting the various advantages of IRS. 
The comparison between the refractive ITS and reflective IRS is illustrated in Fig. \ref{fig:ITSvsIRS}, where an ITS and an IRS are deployed on the $ xy $-plane and $ yz $-plane respectively to aid the communications of a static user equipment (UE) with the base station (BS).  
We thereby propose to embed ITS into the train window to assist the wireless communications along the HSR lines in full length -- Fig.  \ref{fig:system}. 

Hardware implementation of the  ITS is enabled with dielectric metasurfaces 
\cite{ITS_mat_Arbabi_nature,ITS_mat_Feng,ITS_mat_Wu,ITS_mat_Lee},  
which can realize the complete control of transmission amplitude, phase, and polarization. 
Specifically, the metasurface is divided into multiple dielectric unit cells, and one or several meta-atoms made of high-refractive-index amorphous silicon are placed at the centers of these cells.  By properly designing the parameters of meta-atoms, such as geometrical shape (e.g., elliptical, C-shaped, or X-shaped pillar), thickness, orientation, inter-ceil spacing, etc., 
the transmission functionalities of each ITS element can be arbitrarily controlled.  In addition, similar to the IRS, a smart controller can adjust the refraction coefficients of ITS in real time to follow the time-varying wireless channel.  Electronic switches in practice can realize this, 
including diodes, field‐effect transistors (FETs), micro-electromechanical system (MEMS) devices, and so on \cite{ITS_mat_book}.    
Therefore, using the ITS  with HSR communications can ease the penetration loss and cater to the dynamical Doppler shift compensation. 

The channel state information (CSI) is the fundamental knowledge needed for the design of the transceiver and the optimization of the performance of communication networks.  
Note that several works have treated the channel estimation problem for the ITS-aided high-mobility communication systems   
\cite{ITS_Wang,ITS_Lin,
ITS_Huang1,ITS_Huang2,ITS_Wu}, although the ITS is referred to as distinct terminologies. 
Specifically, the reference  \cite{ITS_Lin} first developed the ITS-aided wireless HSR  communication system. This work adopts the ITS and orthogonal time frequency space (OTFS) modulation to overcome the hindrances of high mobility. The joint estimator of channels and Doppler-induced phase shifts for each propagation path is proposed based on the majorization-minimization (MM) method.   
The work \cite{ITS_Wang} studies the joint effective channel estimation and information data detection problem with ITS deployed on the train.
Moreover, the proposed algorithm is evaluated via theoretical analysis.  
The work \cite{ITS_Huang1} considers the   ITS-assisted rural communication system with ITS attached on the top of a high-speed vehicle, and develops a two-stage transmission protocol to execute the channel estimation and beamforming design for data transmission, respectively.  
Therein, both the channels for the link BS-ITS and ITS-UE are assumed to be line-of-sights (LoSs) and then formulated as the functions of channel parameters to be estimated, while the channel for the direct BS-UE link is fast fading owing to various propagation paths. 
We note that the authors in  \cite{ITS_Huang1} extend their work into the general case with the Rician fading (i.e.,  LoS-dominant) channel for the BS-ITS link and reconsider the uniform planar array (UPA) structure at ITS instead of the previous assumption uniform linear array (ULA) \cite{ITS_Huang2}. 
Furthermore, the work \cite{ITS_Wu} simplifies \cite{ITS_Huang1} by considering the direct BS-UE channel as LoS and thus proposes the enhanced transmission protocol with less pilot overhead. 
However, the remaining issues of these works     
\cite{ITS_Huang1,ITS_Huang2,ITS_Wu} 
are presented in the following:

\begin{figure}[t]
\centering
\includegraphics[width=1\linewidth]{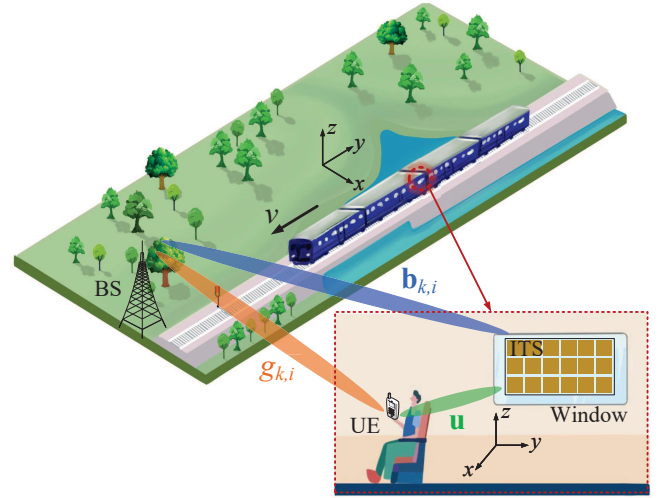}
\caption{An ITS-aided HSR wireless communication system paradigm.}
\label{fig:system}
\end{figure}

\begin{itemize}
\item 
These three works exploit the two-dimension (2D) grid-based search once or more to estimate the channel parameters, and the estimates are further refined via gradient-based search in \cite{ITS_Huang2}. 
Since the performance of the exhaustive search method is directly proportional to the searching precision, 
the desirable performance will be inevitably accompanied by considerable estimation complexity and time consumption.    
\item 
In \cite{ITS_Huang2}, the direct BS-UE  and cascaded BS-ITS-UE channels for each block are repeatedly estimated even in the data transmission stage owing to their time-varying properties, which results in extra training sequence consumption and transmission efficiency loss.  
\item
In \cite{ITS_Wu}, 
the direct channel for each training block is obtained at first by switching off ITS, and next, the cascaded channel can be recovered from the estimate of the direct one with the ITS switched on. 
This method not only enlarges the implementation complexity for the continuous switchover but also degrades the estimation performances of channel parameters for the cascaded link.
In addition, the full potential of beamforming gain can only be achieved after sufficient training blocks for the accurate acquisitions of channel parameters.   
\end{itemize}
 
Motivated by the above, 
we study the design problems of the frame structure and channel estimators for the ITS-assisted HSR communication system (Fig. \ref{fig:system}). 
Since the LoS component is dominant in  HSR wireless because of  the limited   multipath scattering and signal  diffraction from  obstacles   
\cite{HSR_survey/LoS_Ai,HSR_LoS_Wang}, 
we assume that the BS-UE channel is LoS, which is the same assumption made in  \cite{ITS_Wu}.
As such, based on the system model in \cite{ITS_Huang1}, we formulate the  channels with different physical parameters, and the channel estimation is transformed into a parameter recovery  problem. 
In addition, only the first two blocks within each frame are required for pilot transmission, where we keep ITS on and accomplish the entire recovery issue.
In the follow-up blocks, the according channels can be reconstructed via the estimates of channel parameters.
Furthermore, we develop a serial low-complexity channel estimation algorithm by leveraging the classical least square (LS) mechanism. 
  
The main contributions of this paper are summarized
as follows:
\begin{itemize}
\item 
We propose a new two-phase transmission scheme with the ITS consistently turned on. In each frame, only two pilot blocks suffice for the complete channel estimation,   
and the subsequent blocks are all used for data transmission with beamforming performed at the ITS.
To facilitate the channel estimation, every pilot block is further divided into subbloks with each containing training  sequence of identical length.   
\item 
With pre-designed refraction coefficients of  the ITS, we propose a serial low-complexity LS channel estimation algorithm.  Herein, we first precisely obtain the channel  estimates for the direct BS-UE link and the cascaded BS-ITS-UE link,  and formulate each channel estimate as the sum of the perfectly known value and its estimation error. By utilizing the relation between channels for the two pilot blocks, the Doppler shifts can be recovered, based on which we can further estimate  other channel parameters. 
\item 
We derive  the Cram\'er-Rao lower bound (CRLB) for each channel parameter to appraise the performances of the proposed estimators.  Numerical results indicate that our estimators approach or even attain their corresponding CRLBs. 
\end{itemize}


The remainder of this paper is organized as follows. 
Section \ref{System Model} presents the transmission scheme, introduces the system model, and identifies the channel parameters to be estimated. 
Section \ref{CE} proposes the coefficient design of the  ITS and  LS-based channel estimation algorithm.  
Useful discussions can be found in the same section. 
In Section \ref{CRLB}, we derive the CRLBs for our estimators.  
We provide numerical results to corroborate our studies in Section \ref{simulation}. 
Finally, this paper is concluded in Section \ref{conclusion}. 

\textit{Notations:}
Scalars, column vectors and matrices are denoted by italic letters, bold-face lower-case letters, bold-face upper-case letters, respectively. 
Upper-case Calligraphy letters (e.g.,  $\mathcal{K}$) denote discrete and finite sets.    
$ \mathbb{R}^{M\times N} $ and $ \mathbb{C}^{M\times N} $ denote the domain of $ M\times N $ real and complex matrices,  respectively. 
The transpose, conjugate transpose, inverse,   pseudo-inverse, trace and determinant of  $ \mathbf{A} $ are denoted by $ \mathbf{A}^T $, $ \mathbf{A}^H $, $ \mathbf{A}^{-1} $, $ \mathbf{A}^{\dagger} $, $ \operatorname{tr}(\mathbf{A}) $ and $ \operatorname{det}(\mathbf{A}) $, respectively,  and $ [\mathbf{A}]_{m,n} $ denotes the $ (m,n) $th entry.  
For  complex-valued  $ \mathbf{x} $, 
$ \mathcal{R}(\mathbf{x}) $, $ |\mathbf{x}| $ and  $ \angle\mathbf{x} $ denote the vectors with each entry being the real part, amplitude and argument of the corresponding entry in $ \mathbf{x} $, respectively, and $ j $ denotes the imaginary unit, i.e., $ j^2=-1 $. 
For any general vector $ \mathbf{x} $, 
$ [\mathbf{x}]_m $ denotes the $ m $-th entry, 
$ \operatorname{diag}(\mathbf{x}) $ denotes a square diagonal matrix with the elements of $ \mathbf{x} $ on the main diagonal, and  
$ \|\mathbf{x}\|_{\ell} $ denotes the $ \ell $-norm.  
$ \mathbf{1}_{M\times N} $, $ \mathbf{0}_{M\times N} $ and $ \mathbf{I}_{M} $ denote an all-one matrix of size $ M\times N $, an all-zero matrix of size $ M\times N $ and the identity matrix of size $ M\times M $, respectively. 
$ \odot $ and $ \otimes $ denote the Hadamard and Kronecker product, respectively, 
and $ \mathbb{E}_{\mathbf{x}}\{\cdot\} $ denotes the statistical expectation over   random  $ \mathbf{x} $.  
$ \mathcal{CN}(\bm{\mu},\mathbf{\Sigma}) $  denotes the 
distribution of a complex Gaussian random vector with mean $ \bm{\mu} $ and covariance matrix $ \mathbf{\Sigma} $.
$ \sim $ and $ \triangleq $ stand for 
``distributed as" and ``defined as", respectively. 
Finally, ``independently and identically distributed" is abbreviated as ``i.i.d.".

\section{Problem Formulation} \label{System Model}

As shown in Fig. \ref{fig:system}, 
we consider a downlink HSR wireless communication system,
where a UE\footnote{
Without loss of generality, the results of this work can be also applied to the general case with multiple UEs if the time division multiple access (TDMA) technique is exploited.} 
inside the carriage is connected to the BS with the assistance of an ITS embedded into the carriage window.
The train is moving at the high speed of $ v $, 
and the BS stays static.
We assume that both of the BS and UE are equipped with one antenna\footnote{ 
As for the general CSI demand with  multiple antennas employed at BS, 
each of the antennas can be turned on in succession within every two enhanced pilot blocks to individually perform channel estimation. }. 
Moreover, the ITS positioned on the $ yz $-plane is considered to be a UPA  consisting of  $ M=M_y\times M_z $ 
refraction elements, whose amplitudes and phase shifts can be individually controlled in real time by the BS through a smart controller. 

\begin{figure}[t]
\centering
\includegraphics[width=1\linewidth]{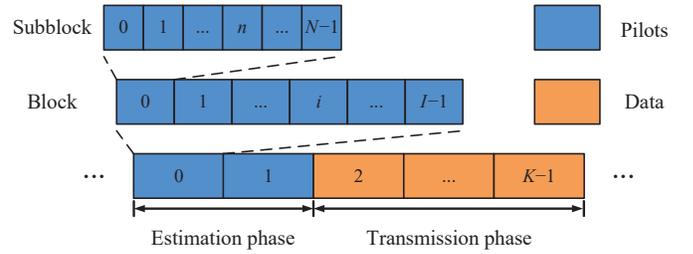}
\caption{Illustration of the proposed two-phase transmission scheme with 
$ N\geq2 $ and $ I\geq\frac{M}{2} $.}
\label{fig:frame}
\end{figure}

A two-phase transmission scheme is proposed in this paper for the ITS-aided HSR communication system (Fig.  \ref{fig:frame}). 
Specifically, each frame with $ K $ blocks is divided into the estimation phase and transmission phase, comprising $ 2 $ pilot blocks and $ K-2 $ data blocks, respectively. 
Moreover, every pilot block is further composed of $ I $ subblocks, with each comprising  $ N $ training pilots.
The minimum quantity of pilot symbols within each frame will be discussed later in Section \ref{CE}.  This current work focuses on the design of the channel estimation algorithm, and the optimization problem of the ITS refraction coefficients in the transmission phase will be forthcoming  in our future work. 
Therefore, we only concentrate on the first two pilot blocks of a certain frame in the remainder of this paper.

\subsection{Channel Model}

Because of the high-speed relative mobility between the BS and train,  both of the link BS-UE and link BS-ITS suffer from severe Doppler shifts.  
Denote the baseband equivalent channels for these two links at the $ i $-th subblock in block $ k $ by 
$ g_{k,i}$ and 
$ \mathbf{b}_{k,i} \in \mathbb{C}^{M\times1} $,  respectively, 
$ k\in \mathcal{K} \triangleq\{0, 1\} $ and 
$ i\in \mathcal{I} \triangleq\{0, \ldots, I-1\} $.
Let $ \mathbf{u} \in \mathbb{C}^{M\times1} $  
represent the equivalent baseband channel for the ITS-UE link,  which stays unchanged within the duration of each  frame, as the ITS and UE are assumed to be relatively static. 
Like \cite{ITS_Wu}, for simplicity, we assume that the three channels mentioned above are all LoS types.  
In this case, the channel $ g_{k,i}$ through the train carriage can be expressed as
\begin{equation}\label{channel_BU}
g_{k,i}=
e^{j 2\pi f_{\textrm{d}1} (kI+i)T }
\rho_{{\textrm{BU}}} \ell_{\textrm{c}},
\end{equation}
where 
$ f_{\textrm{d}1}= \frac{v} {\lambda}
\sin\vartheta^{\textrm{r}}_{\textrm{U}} 
\sin\delta^{\textrm{r}}_{\textrm{U}} $ 
denotes the Doppler frequency shift on the BS-UE link, 
wherein  
$ \lambda $ is the signal wavelength, and  
$ \vartheta^{\textrm{r}}_{\textrm{U}}
(\delta^{\textrm{r}}_{\textrm{U}}) $
denotes the azimuth (elevation) angle of arrival (AoA) at the UE. Moreover, $ T $ represents the duration of each subblock,   the complex-valued parameter  
$ \rho_{\textrm{BU}}$  is the propagation gain for the link BS-UE,   and $ \ell_{\textrm{c}} $ denotes   
the complex penetration gain through the carriage,  which is assumed be a  constant. 
For the convenience of exposition, we define
$ \beta_1 \triangleq \rho_{\textrm{BU}} \ell_{\textrm{c}} $  as the path gain for the direct BS-UE link. 
In addition, the channel $ \mathbf{b}_{k,i} $ can be modelled as 
\begin{equation}\label{channel_BI}
\mathbf{b}_{k,i}=
e^{j 2\pi f_{\textrm{d}2} (kI+i)T}
\rho_{\textrm{BI}}
\underbrace{
\mathbf{a}_y(\varphi_y^\textrm{r}) \otimes
\mathbf{a}_z(\varphi_z^\textrm{r})}
_{\mathbf{a}^{\textrm{r}}
(\vartheta^{\textrm{r}}_{\textrm{T}},\delta^{\textrm{r}}_{\textrm{T}})},
\end{equation}
where 
$ f_{\textrm{d}2}= \frac{v}{\lambda} 
\sin\vartheta^{\textrm{r}}_{\textrm{T}} 
\sin\delta^{\textrm{r}}_{\textrm{T}}  $ 
is the Doppler shift on the  BS-ITS link  with   
$ \vartheta^{\textrm{r}}_{\textrm{T}}
(\delta^{\textrm{r}}_{\textrm{T}}) $
being the azimuth (elevation) AoA at the ITS, 
$ \rho_{\textrm{BI}}$ is the complex propagation gain between the BS and ITS. 
The vector $ {\mathbf{a}^{\textrm{r}}
(\vartheta^{\textrm{r}}_{\textrm{T}},\delta^{\textrm{r}}_{\textrm{T}})} $
represents the receive array response of the ITS in the $ yz $-plane,
$ \mathbf{a}_y(\varphi_y^\textrm{r})\triangleq
\left[1, e^{j\varphi_y^\textrm{r}}, \ldots, e^{j(M_y-1) \varphi_y^\textrm{r}}\right]^T $ and 
$ \mathbf{a}_z(\varphi_z^\textrm{r})\triangleq
\left[1, e^{j \varphi_z^\textrm{r}}, \ldots, e^{j(M_z-1) \varphi_z^\textrm{r}}\right]^T $
are the one-dimension (1D) steering vectors on the $ y $-axis and $ z $-axis, respectively, 
$ \varphi_y^\textrm{r}
\triangleq 
\frac{2\pi}{\lambda} d
\sin\vartheta^\textrm{r}_{\textrm{T}}
\sin\delta^\textrm{r}_{\textrm{T}} $ and 
$ \varphi_z^\textrm{r}
\triangleq 
\frac{2\pi}{\lambda} d
\cos\delta^\textrm{r}_{\textrm{T}} $ 
represent the phase differences between two adjacent elements,  and $ d $ denotes the inter-element spacing, which is equal to half-wavelength. 
We assume that the moving distance of train during one transmission frame is negligible compared to the distance between the BS and train,  and hence the geometry-related channel parameters 
$ \{ f_{\textrm{d}1}, \rho_{\textrm{BU}},
f_{\textrm{d}2}, \rho_{\textrm{BI}},
\vartheta^{\textrm{r}}_{\textrm{T}},
\delta^{\textrm{r}}_{\textrm{T}} \} $
in (\ref{channel_BU}) and (\ref{channel_BI}) can be approximately treated as constants within the considered frame \cite{ITS_Huang1,ITS_Huang2}. 
Similarly, the channel $ \mathbf{u} $ is presented as
\begin{equation}
\mathbf{u}=
\rho_{\textrm{IU}}
\underbrace{
\mathbf{a}_y(\varphi_y^\textrm{t}) \otimes
\mathbf{a}_z(\varphi_z^\textrm{t})}
_{\mathbf{a}^{\textrm{t}}
(\vartheta^{\textrm{t}}_{\textrm{T}},\delta^{\textrm{t}}_{\textrm{T}})},
\end{equation}
where $ \rho_{\textrm{IU}} $ denotes the complex propagation gain for the ITS-UE link, and 
the transmit array response of the ITS
$ {\mathbf{a}^{\textrm{t}}
(\vartheta^{\textrm{t}}_{\textrm{T}},\delta^{\textrm{t}}_{\textrm{T}})} $ 
at an azimuth (elevation) angle of departure (AoD)  
$ \vartheta^{\textrm{t}}_{\textrm{T}}
(\delta^{\textrm{t}}_{\textrm{T}}) $ 
is defined in the same way as the receive array response 
$ {\mathbf{a}^{\textrm{r}}
(\vartheta^{\textrm{r}}_{\textrm{T}},\delta^{\textrm{r}}_{\textrm{T}})} $. 
Herein, the Doppler shift vanishes owing to the relatively montionless assumption as opposed to  (\ref{channel_BU}) and (\ref{channel_BI}).


Let us define   
$ \bm{\phi}_{k,i}= 
\big[ \phi_{k,i}^{(0)}, \ldots, \phi_{k,i}^{(M-1)} \big]
\in\mathbb{C}^{M\times 1} $
as the refraction vector of ITS at the $ i $-th subblock in the $ k $-th block, and 
we tune the amplitude coefficient as 
$ \big| \phi_{k,i}^{(m)} \big|=1 $ for all $ m\in\mathcal{M}\triangleq\{0, \ldots, M-1\} $ 
to eliminate the signal power loss through the ITS.  
We further denote the complex penetration gain through the carriage window by constant   
$ \ell_{\textrm{w}} $.
As such, the effective channel $ \tilde{h}_{k,i} $ for the cascaded BS-ITS-UE link is given by 
\begin{align}\label{channel_BIU}
\tilde{h}_{k,i}
=&\ell_{\textrm{w}}
\big( \mathbf{b}_{k,i} \odot \mathbf{u} \big)^T
\bm{\phi}_{k,i} \nonumber \\
=& e^{j2\pi f_{\textrm{d}2} (kI+i)T} 
\beta_2
\Big( \underbrace{
\mathbf{a}^{\textrm{r}}(\vartheta^{\textrm{r}}_{\textrm{T}},\delta^{\textrm{r}}_{\textrm{T}}) \odot 
\mathbf{a}^{\textrm{t}}(\vartheta^{\textrm{t}}_{\textrm{T}},\delta^{\textrm{t}}_{\textrm{T}}) 
}_{\mathbf{a}} \Big)^T
\bm{\phi}_{k,i},
\end{align}
where 
$ \beta_2\triangleq \rho_{\textrm{BI}}
\rho_{\textrm{IU}}
\ell_{\textrm{w}}   $ 
denotes the path gain for the cascaded link,  
and  $ \mathbf{a} $ is the equivalent array response vector at the ITS, which is expressed as
\begin{align}\label{a}
\mathbf{a}
&=\left(\mathbf{a}_y(\varphi_y^\textrm{r})\otimes
\mathbf{a}_z(\varphi_z^\textrm{r})\right) \odot
\left(\mathbf{a}_y(\varphi_y^\textrm{t}) \otimes
\mathbf{a}_z(\varphi_z^\textrm{t})\right)  \nonumber \\
&\stackrel{(a)}{=}
\left( \mathbf{a}_y(\varphi_y^\textrm{r}) \odot \mathbf{a}_y(\varphi_y^\textrm{t}) \right) \otimes 
\left( \mathbf{a}_z(\varphi_z^\textrm{r}) \odot \mathbf{a}_z(\varphi_z^\textrm{t}) \right) \nonumber \\
&\stackrel{(b)}{=}
\mathbf{a}_y(\varphi_y) \otimes \mathbf{a}_z(\varphi_z),  
\end{align} 
where $ (a) $ is the result of using the mixed property of Kronecker and Hadamard product,  
$ (b) $ is obtained as we perform the Hadamard product and then define 
$ \varphi_y \triangleq  \varphi_y^\textrm{r}+\varphi_y^\textrm{t} $ and   
$ \varphi_z \triangleq  \varphi_z^\textrm{r}+\varphi_z^\textrm{t} $ 
as the equivalent inter-element phase differences on the $ y $ and $ z $ axes, respectively.  

 


\subsection{Signal Model}

Let $ \mathcal{N} \triangleq\{0, \ldots, N-1\} $ 
denote the index set of training symbols in each subblock. We further consider that the refraction vector of the ITS is adjusted symbol by symbol, say 
$ \bm{\phi}_{k,i,n} $, which is designed as   
\begin{equation}\label{phi_kin}
\bm{\phi}_{k,i,n}=
\psi_{k,i,n} \bar{\bm{\phi}}_{k,i},
\end{equation}
where $ \bar{\bm{\phi}}_{k,i} $ denotes the initial refraction vector dedicated to subblock $ i $ of block $ k $, 
and $ \psi_{k,i,n} $ is the adjustment coefficient for pilot $ n $. 

As the transmitted signal from the BS can propagate to the UE through the direct BS-UE link and the cascaded BS-ITS-UE link, 
the received signal at the $ n $-th symbol in the $ i $-th subblock of block $ k $ is given by
\begin{equation}
y_{k,i,n}=
(g_{k,i}+\psi_{k,i,n} h_{k,i}) x_{k,i,n}
+w_{k,i,n},
\end{equation}
where 
$ x_{k,i,n} $ denotes the training symbol which is simply designed as $ x_{k,i,n}=1 $ for the exposition simplicity, 
$ w_{k,i,n}\sim\mathcal{CN}(0,\sigma^2) $
is the additive i.i.d. zero-mean complex Gaussian noise at the UE, 
and $ h_{k,i} $ represents the initial cascaded channel at each subblock, i.e., 
\begin{equation}
h_{k,i}=
e^{j2\pi  f_{\textrm{d}2} (kI+i)T } 
\beta_2 \mathbf{a}^T \bar{\bm{\phi}}_{k,i}.
\end{equation}



\begin{remark}
Our target now is to estimate the unknown channel parameters, including: (i) Doppler shifts 
$ \{f_{\textrm{d}1},f_{\textrm{d}2} \} $, 
(ii) path gains 
$ \{ \beta_{1},\beta_{2} \} $, 
and (iii) equivalent phase differences at ITS 
$ \{ \varphi_y,\varphi_z \} $.
\end{remark}

\section{Channel Estimation Algorithm}\label{CE}

\subsection{Channel Estimation}
We start by tackling the estimation problem of channels 
$ g_{k,i} $ and $ h_{k,i} $ with $ k\in\mathcal{K} $, $ i\in\mathcal{I} $.
To facilitate the channel estimation,   
with the aid of (\ref{phi_kin}), 
the refraction coefficients of ITS are pre-designed by letting    
$ \psi_{0,i,n}=\psi_{1,i,n} $ and 
$ \bar{\bm{\phi}}_{0,i}=\bar{\bm{\phi}}_{1,i} $ 
for all $ i $ and $ n $, 
and thus we have 
$ \bm{\phi}_{0,i,n}=\bm{\phi}_{1,i,n} $. 
In this case, for notation simplicity, we omit their subscript $ k $  in the sequel. 
As a result, by stacking $ 2N $ received pilot symbols 
$ \{ y_{0,i,n} \}_{n=0}^{N-1} $ and 
$ \{ y_{1,i,n} \}_{n=0}^{N-1} $ 
at the $ i $-th subblock of both block 0 and block 1, 
the received signal can be expressed in the vector form as 
\begin{equation}\label{Yi}
\mathbf{Y}_i=
\mathbf{\Psi}_i \mathbf{V}_i + \mathbf{W}_i,
\end{equation}
where 
$ \mathbf{Y}_i= 
\left[ \mathbf{y}_{0,i},\mathbf{y}_{1,i} \right]
\in\mathbb{C}^{N\times2} $ 
denotes the received symbol matrix with
$ \mathbf{y}_{k,i}= 
\left[ y_{k,i,0},\ldots,y_{k,i,N-1} \right]^T
\in\mathbb{C}^{N\times1} $, 
$ \mathbf{\Psi}_i 
= \left[ \mathbf{1}_{N\times1},
\bm{\psi}_{i} \right]  \in\mathbb{C}^{N\times2} $ 
represents the training matrix with 
$ \bm{\psi}_{i} 
= \left[ \psi_{i,0},\ldots, \psi_{i,N-1} \right]^T \in\mathbb{C}^{N\times1} $,  
$ \mathbf{V}_i=
\left[ \mathbf{v}_{0,i}, \mathbf{v}_{1,i} \right]
\in\mathbb{C}^{2\times2}
 $ 
is the channel matrix to be estimated with  
$ \mathbf{v}_{k,i}=
\left[ g_{k,i}, h_{k,i} \right]^T $,
and denote the received noise matrix by 
$ \mathbf{W}_i=
\left[ \mathbf{w}_{0,i},\mathbf{w}_{1,i} \right]
\in\mathbb{C}^{N\times2} $ with 
$ \mathbf{w}_{k,i}= 
\left[ w_{k,i,0},\ldots,w_{k,i,N-1} \right]^T
\in\mathbb{C}^{N\times1} $. 

By proper design of the training matrix $ \mathbf{\Psi}_{i}  $,  
we can construct a full-rank tall rectangular matrix, namely $ \mathbf{\Psi}_{i}^{\dagger} $ exists.  
In this paper, we exploit the first and another columns of the $ N $-point discrete Fourier transform (DFT) matrix with
$ \left[\mathbf{F}_N\right]_{t_1,t_2}=
e^{-j\frac{2\pi}{N} t_1t_2} $ for  
$ 0\leq t_1,t_2 \leq N-1 $, 
therefore 
$ \mathbf{\Psi}_i^H \mathbf{\Psi}_i=N \mathbf{I}_2,
\forall i\in\mathcal{I} $.
Under this condition, 
the estimate of channel matrix for the $ i $-th subblocks in the first two blocks can be obtained via the LS estimation. That is, we have  
\begin{align}\label{channel_estimate}
\widehat{\mathbf{V}}_i&=
(\mathbf{\Psi}_i^H \mathbf{\Psi}_i)^{-1} \mathbf{\Psi}_i^H 
\mathbf{Y}_i \nonumber \\ 
&= \mathbf{V}_i  
+ \frac{1}{N} \mathbf{\Psi}_i^H \mathbf{W}_i. 
\end{align}

At the end of the estimation phase, we can reap 
$ \hat{g}_{k,i} $ and $ \hat{h}_{k,i} $ for every $ k $ and $ i $, 
which will be further utilized to acquire estimates of the unknown channel parameters. The details of this process are  in the following two subsections.


\subsection{Doppler Shift Estimation}
\subsubsection{Direct Link}
Let us formulate the acquired channel estimate for the direct BS-UE link $ \hat{g}_{k,i} $ as the sum of $ g_{k,i} $ and its estimation error $ \varepsilon_{k,i} $, and collect 
$ \{ \hat{g}_{k,i} \}_{i=0}^{I-1} $   
within each pilot block respectively, yielding 
\begin{align}
\label{g0}
\widehat{\mathbf{g}}_{0}&=
\beta_1 \mathbf{d}_1 + \bm{\varepsilon}_0, \\
\label{g1}
\widehat{\mathbf{g}}_{1}&=
\beta_1 \xi_1\mathbf{d}_1 + \bm{\varepsilon}_1,
\end{align}
where
$ \bm{\varepsilon}_k=
\left[ \varepsilon_{k,0},\ldots,\varepsilon_{k,I-1} \right]^T
\in\mathbb{C}^{I\times1} $ 
represents the estimation error vector of  
$ \mathbf{g}_{k}=
\left[ g_{k,0},\ldots,g_{k,I-1} \right]^T $,  
and note that  
$ \varepsilon_{k,i}=
\frac{1}{N} \mathbf{1}_{N\times1}^T \mathbf{w}_{k,i}
\sim \mathcal{CN} (0,\frac{\sigma^2}{N}) $ 
based on (\ref{channel_estimate}), 
$ \forall k\in\mathcal{K} $ and $ \forall i\in\mathcal{I} $, 
$ \mathbf{d}_{1}=
\left[ 
1, e^{j2\pi f_{\textrm{d}1}T} ,\ldots, 
e^{j2\pi f_{\textrm{d}1} (I-1)T }  \right]^T
\in\mathbb{C}^{I\times1} $ 
is the Doppler-induced phase shift vector for the link BS-UE in block 0,   
$ \xi_1 \triangleq e^{j 2\pi f_{\textrm{d}1} IT} $ 
denotes the phase shift gap between the $ i $-th subblocks of block 0 and block 1 for all $ i $, 
and we let 
$ \alpha_1\triangleq 2\pi f_{\textrm{d}1} IT $ 
for ease of exposition. 

Importantly, the variance of every $ \varepsilon_{k,i} $ degrades into $ \frac{1}{N} $ noise power, and the estimation error of each $ h_{k,i} $, say $ \varsigma_{k,i} $, possesses the same  variance above (to be presented soon). This is favorable for the more precise acquisitions of   channel parameter estimates in the following process.

\begin{remark}
Although the growth of $ N $ is beneficial for the channel estimates as presented above, the sample amount of which in one block (i.e., $ I $ for each link) is on the decline at the same time with training sequence of fixed length. 
To reap the optimal estimation performances of channel parameters, 
it is therefore expected to consider the design trade-off between $ N $ and $ I $.  
\end{remark}

In order to eliminate the unknown $ \beta_1 $, substitute (\ref{g0}) into (\ref{g1}), 
and $ \widehat{\mathbf{g}}_{1} $ can be rewritten as 
\begin{equation}\label{g1g0}
\widehat{\mathbf{g}}_{1}=
\xi_1 \big( \widehat{\mathbf{g}}_{0} - \bm{\varepsilon}_0 \big)
+\bm{\varepsilon}_1 
=\xi_1\widehat{\mathbf{g}}_{0} +\widetilde{\bm{\varepsilon}},
\end{equation} 
where
$ \widetilde{\bm{\varepsilon}} 
=\bm{\varepsilon}_1 -\xi_1\bm{\varepsilon}_0 $
denotes the effective estimation error vector,  
and 
$ \widetilde{\bm{\varepsilon}}  \sim
\mathcal{CN} (0,\frac{2\sigma^2}{N} \mathbf{I}_I) $. 
Since $ |\xi_1|=1 $, $ \hat{\xi}_1 $ can be obtained via the normalized LS estimator as  
\begin{equation}
\hat{\xi}_1=
\big\| \widehat{\mathbf{g}}_{0}^H \widehat{\mathbf{g}}_{1} \big\|_2^{-1}
\widehat{\mathbf{g}}_{0}^H 
\widehat{\mathbf{g}}_{1}, 
\end{equation}
where we have exploited that the vector product 
$ \widehat{\mathbf{g}}_0^H \widehat{\mathbf{g}}_0 $ is real. 
Once $ \hat{\xi}_1 $ is obtained, 
$ \hat{f}_{\textrm{d}1} $ can be found from
\begin{equation}\label{fd1}
\hat{f}_{\textrm{d}1}=
\frac{\angle\hat{\xi}_1}{2\pi IT}.
\end{equation}

\subsubsection{Cascaded Link}
To get  the estimates of channel parameters for the cascaded BS-ITS-UE link, we similarly aggregate 
$ \{ \hat{h}_{k,i} \}_{i=0}^{I-1} $ for each   
$ k\in\mathcal{K} $, i.e., 
\begin{align}
\label{h0}
\widehat{\mathbf{h}}_{0}&=
\beta_2 \mathbf{D}_2 \bar{\mathbf{\Phi}} \mathbf{a} + \bm{\varsigma}_0, \\
\label{h1}
\widehat{\mathbf{h}}_{1}&=
\beta_2 \xi_2 \mathbf{D}_2 \bar{\mathbf{\Phi}} \mathbf{a} + \bm{\varsigma}_1,
\end{align}
where
$ \bm{\varsigma}_k=
\left[ \varsigma_{k,0}, \ldots, \varsigma_{k,I-1} \right]^T
\in\mathbb{C}^{I\times1} $ 
is the estimation error vector corresponding to  
$ \mathbf{h}_{k}=
\left[ h_{k,0},\ldots,h_{k,I-1} \right]^T $,  
and each error is given by    
$ \varsigma_{k,i}=
\frac{1}{N} \bm{\psi}_i^H \mathbf{w}_{k,i} $ for all $ k $ and $ i $   
whose distribution is identical with $ \varepsilon_{k,i} $,  
i.e, $ \mathcal{CN} (0,\frac{\sigma^2}{N}) $, 
$ \mathbf{D}_{2}=
\operatorname{diag} \!\left\{ 
1, e^{j2\pi f_{\textrm{d}2} T} ,\ldots,
e^{j2\pi f_{\textrm{d}2} (I-1)T}  \right\}
\in\mathbb{C}^{I \times I} $ 
denotes the phase shift diagonal matrix for the cascaded link owing to the Doppler effect, 
$ \xi_2 \triangleq 
e^{j \alpha_2} $ with its argument 
$ \alpha_2\triangleq 2\pi f_{\textrm{d}2} IT $, and 
$ \bar{\mathbf{\Phi}}=
\left[ \bar{\bm{\phi}}_0,\ldots, 
\bar{\bm{\phi}}_{I-1} \right]^T
\in\mathbb{C}^{I\times M} $ 
is the initial refraction matrix of ITS which can be designed in advance 
(shown in the next subsection). 

With (\ref{h0}),
$ \widehat{\mathbf{h}}_{1} $ 
in (\ref{h1}) can be recast as
\begin{equation}\label{h1h0}
\widehat{\mathbf{h}}_{1}=
\xi_2 \big( \widehat{\mathbf{h}}_{0} 
-\bm{\varsigma}_0 \big)
+\bm{\varsigma}_1 
=\xi_2 \widehat{\mathbf{h}}_{0} +\widetilde{\bm{\varsigma}},
\end{equation} 
where
$ \widetilde{\bm{\varsigma}}=
\bm{\varsigma}_1 -\xi_2\bm{\varsigma}_0 $ 
is distributed as 
$ \mathcal{CN}
(0,\frac{2\sigma^2}{N} \mathbf{I}_I)  $. 
As such, the normalized LS estimator\footnote{
Intuitively, with the prior knowledge 
$ |\xi_1|,|\xi_2|=1 $, 
each normalized LS estimator outperforms the conventional one. 
This insight will be verified by simulation in Section \ref{simulation}. }  of $ \xi_2 $ is expressed as 
\begin{equation}
\hat{\xi}_2
=\big\| \widehat{\mathbf{h}}_{0}^H
\widehat{\mathbf{h}}_{1} \big\|_2^{-1}
\widehat{\mathbf{h}}_{0}^H
\widehat{\mathbf{h}}_{1}.
\end{equation}
Accordingly, 
$ \hat{f}_{\textrm{d}2} $ is given by 
\begin{equation}\label{fd2}
\hat{f}_{\textrm{d}2}=
\frac{\angle\hat{\xi}_2}{2\pi IT}.
\end{equation} 

\begin{remark}
Note that the complex number is a periodic function of its argument with the period $ 2\pi $, thus it is widely applied that the argument will be converted into the range 
$ \left( -\pi,\pi \right] $ 
when dealing with the operator $ \angle(\cdot) $.  
In other words, the argument we obtain is the result of modulo $ 2\pi $. 
Under this condition, to enable the Doppler shift estimators (\ref{fd1}) and (\ref{fd2}), 
we can appropriately set the subblock quantity within one block $ I $ and the sublock duration $ T  $ to ensure 
$ \alpha_1,\alpha_2 \in \left( -\pi,\pi \right] $.
\end{remark}

\subsection{Path Gain and Phase Difference Estimation }

\subsubsection{Direct Link}
With the aid of $ \hat{f}_{\textrm{d}1} $, we can construct the estimate of vector $ \mathbf{d}_1 $.  Consequently, based on (\ref{g0}), the LS estimator of $ \beta_1 $ is expressed as
\begin{equation}
\hat{\beta}_{1}= 
\frac{1}{I}\,
\widehat{\mathbf{d}}_1^H \widehat{\mathbf{g}}_0, 
\end{equation} 
since we have realized that  
$ \widehat{\mathbf{d}}_1^H 
\widehat{\mathbf{d}}_1=I $.

\subsubsection{Cascaded Link}
Given (\ref{h0}) and (\ref{h1}), 
we first consider the vertical concatenation of 
$ \widehat{\mathbf{h}}_{0} $ and 
$ \widehat{\mathbf{h}}_{1} $, i.e.,   
\begin{equation}\label{h}
\widehat{\mathbf{h}}=
\beta_2 \mathbf{\Gamma} \widetilde{\mathbf{\Phi}} \mathbf{a} + \bm{\varsigma},
\end{equation}
where 
$ \widehat{\mathbf{h}}= 
\left[ \widehat{\mathbf{h}}_0^T,
\widehat{\mathbf{h}}_1^T \right]^T 
\in\mathbb{C}^{2I\times1},   
\bm{\varsigma}= 
\left[ \bm{\varsigma}_0^T, 
\bm{\varsigma}_1^T \right]^T $,
$ \mathbf{\Gamma}= $
$\left[ \begin{matrix}
\mathbf{D}_2 &\mathbf{0}_{I\times I}\\ 
\mathbf{0}_{I\times I} &\xi_2 \mathbf{D}_2
\end{matrix} \right]
\in\mathbb{C}^{2I \times 2I} $ and 
$ \widetilde{\mathbf{\Phi}}=
\left[ \begin{matrix}
\,\bar{\mathbf{\Phi}}\, \\ \,\bar{\mathbf{\Phi}}\,
\end{matrix} \right]
\in\mathbb{C}^{2I\times M} $ 
denote the Doppler-induced phase shift diagonal matrix and the initial ITS refraction matrix corresponding to the first two blocks, respectively. 

Based on 
$ \hat{f}_{\textrm{d}1} $,   
we can obtain the estimate of $ \mathbf{\Gamma} $.  
As for the design method of 
$ \widetilde{\mathbf{\Phi}} $, 
similar to  
$ \mathbf{\Psi}_i,\forall i\in\mathcal{I} $ in (\ref{Yi}), 
we can preset $ \bar{\mathbf{\Phi}} $ as the first $ M $ columns of the $ I $-th point DFT matrix 
$ \mathbf{F}_I $ with $ I>M $, 
and $ \widetilde{\mathbf{\Phi}} $ is the vertical combination of $ \bar{\mathbf{\Phi}} $ and its duplicate. 
In this case, it is evident that the matrix product 
$ \widehat{\mathbf{\Gamma}}\widetilde{\mathbf{\Phi}} $ is a full-rank tall rectangular matrix, and we have such that 
$ \big(
\widehat{\mathbf{\Gamma}}\widetilde{\mathbf{\Phi}} \big)^H
\widehat{\mathbf{\Gamma}}\widetilde{\mathbf{\Phi}}
=2I\mathbf{I}_M $.
Moreover, define  
$ \mathbf{c} \triangleq \beta_2\mathbf{a} $. 
According to $ (b) $ in (\ref{a}), 
further conduct the Kronecker product and let 
$ \beta_2=|\beta_2|e^{j \angle \beta_2} $, then 
$ \mathbf{c} $ can be rewritten as 
\begin{equation}
\mathbf{c}=|\beta_2|e^{j\mathbf{\Omega}\mathbf{z}}, 
\end{equation}
where 
$ \mathbf{z}=\left[ 
\angle\beta_2, \varphi_y, \varphi_z \right]^T $, and    
$ \mathbf{\Omega}
=\left[ \mathbf{1}_{M\times1}, 
\bm{\varpi}_y, \bm{\varpi}_z \right]
 \in \mathbb{R}^{M\times3} $ 
denotes the coefficient matrix with the last two columns defined as     
$ \bm{\varpi}_y=
\left[0,\ldots,M_{y}-1\right]^T \otimes\mathbf{1}_{M_z\times1} $ and 
$ \bm{\varpi}_z= 
\mathbf{1}_{M_y\times1} \otimes \left[0,\ldots,M_{z}-1\right]^T $, respectively.   
This way, we can easily verify the existence of 
$ \mathbf{\Omega}^{\dagger} $ 
when $ M_y\geq2 $ and $ M_z\geq2 $. 

With (\ref{h}), the LS estimate 
$ \widehat{\mathbf{c}} $ is given by 
\begin{equation}\label{c}
\widehat{\mathbf{c}}= \frac{1}{2I}
\widetilde{\mathbf{\Phi}}^H 
\widehat{\mathbf{\Gamma}}^H 
\widehat{\mathbf{h}}.
\end{equation}
Therefore, $ |\hat{\beta}_2| $ is expressed as 
\begin{equation}
|\hat{\beta}_2|=
\frac{\left\| \widehat{\mathbf{c}} \right\|_1}{M}. 
\end{equation}
Like $ \alpha_1 $ and $ \alpha_2 $, in order to validate the argument estimation, 
assume that the argument of every entry in $ \mathbf{c} $ belongs to range 
$ \left( -\pi,\pi \right] $. Consequently,   
$ \widehat{\mathbf{z}} $ is found from 
\begin{equation}\label{zhat}
\widehat{\mathbf{z}}=
\left( \mathbf{\Omega}^H \mathbf{\Omega} \right)^{-1} \mathbf{\Omega}^H  \angle\,\widehat{\mathbf{c}}.
\end{equation}

\begin{remark}
Similar to the estimator of $ \beta_2 $, 
we find  that $ \hat{\beta}_1 $ can be obtained via the vertical concatenation of 
$ \widehat{\mathbf{g}}_0 $ and $ \widehat{\mathbf{g}}_1 $. 
Intuitively, this method helps to boost the estimation performance due to more samples of the  channel estimate. 
However, it can be verified that the performance gain of $ \hat{\beta}_1 $ is relatively small, and the gap with respect to its performance bound is enlarged instead.  
\end{remark}

\begin{remark}
According to (\ref{channel_estimate}), at least two pilots in each subblock are required for the estimation of $ \mathbf{V}_i $ for all $ i\in\mathcal{I} $, i.e., $ N\geq2 $. 
Moreover, to activate the estimator of $ \mathbf{c} $ (\ref{c}), the subblock quantity within one block has to satisfy such that $ I\geq\frac{M}{2} $. 
Since we have designed two blocks for pilot transmission, the minimum amount of total training   symbols in each frame becomes $ 2M $, which only depends on the ITS element quantity $ M $. 
Future studies can be carried out to further economize the pilot consumption, such as the ITS  grouping method as the result of the high channel spatial correlation \cite{IRS_group}. 

\end{remark}

\section{Derivation of the CRLBs}\label{CRLB}

Herein, we will derive the CRLBs to  evaluate the performances of the proposed channel parameter estimators. 

\subsection{CLRBs of Doppler Shifts}
Inspired by the fact that  
$ \hat{f}_{\textrm{d}1} (\hat{f}_{\textrm{d}2}) $ 
is acquired via the argument of  
$ \hat{\xi}_1 (\hat{\xi}_2) $, 
we will introduce CRLBs to bound their estimation variances 
respectively in this subsection and compare  performances later through numerical results in Section \ref{simulation}.   

\subsubsection{Direct Link}
We start with the derivation of CRLB for $ \hat{\xi}_1 $ in detail.   
Using (\ref{g1g0}) with given 
$ \{\xi_1,\widehat{\mathbf{g}}_{0}
,\bm{\varepsilon}_0\} $, 
the $ I $-th dimension vector  
$ \widehat{\mathbf{g}}_{1} $ 
is distributed as conditionally complex Gaussian, 
and its likelihood function is written as 
\begin{equation}
p\!\left( \widehat{\mathbf{g}}_{1}\big|
\xi_1,\widehat{\mathbf{g}}_{0},\bm{\varepsilon}_0 \right)=
\frac{1}{\det \! \left(
 \pi\mathbf{C}_{\widehat{\mathbf{g}}_{1}} \!\right)}
e^{- \left( \widehat{\mathbf{g}}_{1}  
-\bm{\mu}_{\widehat{\mathbf{g}}_{1}} \!\right)^H
\mathbf{C}_{\widehat{\mathbf{g}}_{1}}^{-1} 
\left( \widehat{\mathbf{g}}_{1}  
-\bm{\mu}_{\widehat{\mathbf{g}}_{1}} \!\right) },
\end{equation} 
where it is easily shown that 
\begin{equation}\label{mug1}
\bm{\mu}_{\widehat{\mathbf{g}}_{1}}=
\mathbb{E}_{\bm{\varepsilon}_1}\!\!
\left[ \,\widehat{\mathbf{g}}_{1} \right]=
\xi_1 \widehat{\mathbf{g}}_{0} - \xi_1\bm{\varepsilon}_0,
\end{equation}
and that 
\begin{equation}\label{Cg1}
\mathbf{C}_{\widehat{\mathbf{g}}_{1}}=
\mathbb{E}_{\bm{\varepsilon}_1}\!
\big[ \left( \widehat{\mathbf{g}}_{1}  
-\bm{\mu}_{\widehat{\mathbf{g}}_{1}} \!\right)
\left( \widehat{\mathbf{g}}_{1}  
-\bm{\mu}_{\widehat{\mathbf{g}}_{1}} \!\right)^H \!\big]
=\frac{\sigma^2}{N} \mathbf{I}_I.
\end{equation}
Introducing the log-likelihood function 
\begin{align}
\Lambda_{\widehat{\mathbf{g}}_{1}}&=
\ln p\!\left( \widehat{\mathbf{g}}_{1} \big|
\xi_1,\widehat{\mathbf{g}}_{0},\bm{\varepsilon}_0 
      \right) \nonumber \\
&=-\ln\!\left(  \det \! \left(\pi\mathbf{C}_{\widehat{\mathbf{g}}_{1}} \!\right) \right)
-\left( \widehat{\mathbf{g}}_{1}  
-\bm{\mu}_{\widehat{\mathbf{g}}_{1}} \!\right)^H
\mathbf{C}_{\widehat{\mathbf{g}}_{1}}^{-1} 
\left( \widehat{\mathbf{g}}_{1}  
-\bm{\mu}_{\widehat{\mathbf{g}}_{1}} \!\right). 
\end{align}
With (\ref{mug1}) and (\ref{Cg1}), after some tedious manipulations, we further have 
\begin{align}
\Lambda_{\widehat{\mathbf{g}}_{1}}
=-I&\ln\! \Big(\Big.\Big.\frac{\pi\sigma^2}{N}\Big)
-\frac{N}{\sigma^2}
\Big(
\widehat{\mathbf{g}}_{1}^H \widehat{\mathbf{g}}_{1}
+2\mathcal{R} \big\{
 \xi_1  \widehat{\mathbf{g}}_{1}^H \bm{\varepsilon}_0
- \xi_1 \widehat{\mathbf{g}}_{1}^H  \widehat{\mathbf{g}}_{0} \big\}
\bigg.\nonumber \\
&\bigg.+|\xi_1|^2
\big\{ \widehat{\mathbf{g}}_{0}^H \widehat{\mathbf{g}}_{0}
-2\mathcal{R}\{ \bm{\varepsilon}_0^H \widehat{\mathbf{g}}_{0} \}
+\bm{\varepsilon}_0^H \bm{\varepsilon}_0 \big\} \Big).
\end{align}

Since $ \xi_1 $ is complex, we suggest to partition its real part and imaginary part, denoted by 
$ \xi_{1,\textrm{R}} $ and $ \xi_{1,\textrm{I}} $  respectively, and define the real 2D vector  
$ \bm{\xi}_1=
\left[\xi_{1,\textrm{R}}, 
\xi_{1,\textrm{I}} \right]^T $. 
Then, the bound on the variance of 
$ \bm{\xi}_1 $ are given by 
\begin{equation}
\mathbb{E}
\big[ \big(\bm{\xi}_1-\widehat{\bm{\xi}}_1\big) 
\big(\bm{\xi}_1-\widehat{\bm{\xi}}_1 \big)^T \big]
\geq \bm{\mathcal{F}}_{\widehat{\mathbf{g}}_{1}}^{-1} (\bm{\xi}_1),
\end{equation}
where
$ \bm{\mathcal{F}}_{\widehat{\mathbf{g}}_{1}} (\bm{\xi}_1) $   
denotes the Fisher information matrix (FIM) of 
$ \widehat{\mathbf{g}}_{1} $ 
with parameter vector   
$ \bm{\xi}_1 $, 
which is defined as 
\begin{equation}
\bm{\mathcal{F}}_{\widehat{\mathbf{g}}_{1}}(\bm{\xi}_1)
=-\mathbb{E} \left\{ 
\left[\begin{matrix}
\frac{\partial^{2}\Lambda_{\widehat{\mathbf{g}}_{1}}}
{\partial \xi_{1,\textrm{R}}^{2}} 
& \frac{\partial^{2}\Lambda_{\widehat{\mathbf{g}}_{1}}}
{\partial \xi_{1,\textrm{R}} \xi_{1,\textrm{I} } }  \\
\frac{\partial^{2}\Lambda_{\widehat{\mathbf{g}}_{1}}}
{\partial \xi_{1,\textrm{I}}\xi_{1,\textrm{R}} } 
& \frac{\partial^{2}\Lambda_{\widehat{\mathbf{g}}_{1}}}
{\partial \xi_{1,\textrm{I}}^2 }  
\end{matrix}\right]   \right\},
\end{equation}
whose both off-diagonal entries are  zeros, while diagonal entries are  identically given by 
\begin{align}
&\big[
\bm{\mathcal{F}}_
{\widehat{\mathbf{g}}_{1}}  
(\bm{\xi}_1) \big]_{1,1} 
=\big[
\bm{\mathcal{F}}_
{\widehat{\mathbf{g}}_{1}}  
(\bm{\xi}_1) \big]_{2,2}
\nonumber \\
&=\frac{2N}{\sigma^2}
\Big\{ \mathbb{E}\!\left[ \widehat{\mathbf{g}}_{0}^H \widehat{\mathbf{g}}_{0} \right]
-2\mathcal{R}\!\left\{ \mathbb{E}\!\left[ \bm{\varepsilon}_0^H \widehat{\mathbf{g}}_{0} \right] \right\}
+\mathbb{E}\!\left[\bm{\varepsilon}_0^H  \bm{\varepsilon}_0 \right] \! \Big\} 
=\frac{2NI}{\sigma^2},  
\end{align}
where we have assumed that 
$ |\beta_1|=1 $  
(also in the remainder of this paper) for the brevity of exposition, 
and the statistical expectations have been taken with respect to $ \bm{\varepsilon}_0 $.   
Consequently, the CRLB for the estimator of $ \xi_1 $ is expressed as 
\begin{equation}
\mathbb{E}\big[\big| \xi_1 
-\hat{\xi}_1 \big|^2 \big]
\geq \operatorname{tr}\!
\big\{( \bm{\mathcal{F}}_{\widehat{\mathbf{g}}_{1}}^{-1}
(\bm{\xi}_1) \big\}
=\frac{\sigma^2}{NI}.
\end{equation}

Moreover, consider the CRLB for 
$ \hat{f}_{\textrm{d}1} $. 
Since only the parameter $ f_{\textrm{d}1} $ is to be estimated with the prior information 
$ |\xi_1|=1 $, 
the corresponding FIM 
$  \mathcal{F}_{\widehat{\mathbf{g}}_{1}} (f_{\textrm{d}1})  $
degrades to scalar in this case, 
and its element is written as     
\begin{align}
-\mathbb{E}\!\left[
\frac{\partial^{2}\Lambda_{\widehat{\mathbf{g}}_{1}}}
{\partial f_{\textrm{d}1}^2 } \right]
&=\frac{8 \pi^2 N I^2 T^2}{\sigma^2}
\mathcal{R} \Big\{ \xi_1 \big\{  \mathbb{E}\!\left[ 
 \widehat{\mathbf{g}}_{1}^H 
\widehat{\mathbf{g}}_{0}  \right]
-\mathbb{E}\big[
\widehat{\mathbf{g}}_{1}^H 
\bm{\varepsilon}_0  \big] \big\}  \Big\} \nonumber \\
&=\frac{8 \pi^2 N I^3 T^2}{\sigma^2},
\end{align}
from which we have   
\begin{equation}
\mathbb{E}\big[\big| f_{\textrm{d}1}
-\hat{f}_{\textrm{d}1} \big|^2 \big]
\geq \mathcal{F}^{-1}_{\widehat{\mathbf{g}}_{1}} (f_{\textrm{d}1})
=\frac{\sigma^2}{8 \pi^2 N I^3 T^2}.
\end{equation}

\begin{figure*}[hb]
\centering
\normalsize
\setcounter{MYtempeqncnt}{\value{equation}}
\vspace*{4pt}
\hrulefill
\setcounter{equation}{49}
\begin{equation}\label{FIMl}
\bm{\mathcal{F}}_{\widehat{\mathbf{h}}}
(\bar{\mathbf{z}})= \frac{4NI}{\sigma^2}  
\left[\begin{matrix}
M &0 
&-\beta_{2,\textrm{I}}\left\| \bm{\varpi}_y \right\|_1 
&-\beta_{2,\textrm{I}}\left\| \bm{\varpi}_z \right\|_1
\\
0 &M 
&\beta_{2,\textrm{R}}\left\| \bm{\varpi}_y \right\|_1 
&\beta_{2,\textrm{R}}\left\| \bm{\varpi}_z \right\|_1
\\
-\beta_{2,\textrm{I}}\left\| \bm{\varpi}_y \right\|_1 
&\beta_{2,\textrm{R}}\left\| \bm{\varpi}_y \right\|_1
&\left\| \bm{\varpi}_y \right\|_2^2
&\bm{\varpi}_y^T \bm{\varpi}_z
\\
-\beta_{2,\textrm{I}}\left\| \bm{\varpi}_z \right\|_1 
&\beta_{2,\textrm{R}}\left\| \bm{\varpi}_z \right\|_1
&\bm{\varpi}_z^T \bm{\varpi}_y 
&\left\| \bm{\varpi}_z \right\|_2^2
\end{matrix}\right].
\end{equation}
\setcounter{equation}{\value{equation}}
\end{figure*}

\subsubsection{Cascaded Link}

Similarly, let us first address the bounding problem of $ \hat{\xi}_2 $. 
Based on (\ref{h1h0}), 
the likelihood function of  
$ \widehat{\mathbf{h}}_1 $ with perfectly known 
$ \{\xi_2,\widehat{\mathbf{h}}_{0}
,\bm{\varsigma}_0\} $
can be formulated as 
\begin{equation}
\setcounter{equation}{38}
p\big( \widehat{\mathbf{h}}_{1}\big|
\xi_2,\widehat{\mathbf{h}}_{0},\bm{\varsigma}_0 \big)=
\frac{1}{\det \! \big(
 \pi\mathbf{C}_{\widehat{\mathbf{h}}_{1}} \!\big)}
e^{- \big( \widehat{\mathbf{h}}_{1}  
-\bm{\mu}_{\widehat{\mathbf{h}}_{1}} \!\big)^H
\mathbf{C}_{\widehat{\mathbf{h}}_{1}}^{-1} 
\big( \widehat{\mathbf{h}}_{1}  
-\bm{\mu}_{\widehat{\mathbf{h}}_{1}} \!\big) },
\end{equation} 
where
$ \bm{\mu}_{\widehat{\mathbf{h}}_{1}}=
\xi_2\widehat{\mathbf{h}}_{0} 
- \xi_2\bm{\varsigma}_0 $ and 
$ \mathbf{C}_{\widehat{\mathbf{h}}_{1}}=
\frac{\sigma^2}{N} \mathbf{I}_I $. 
Consider the logarithm likelihood function
\begin{align}
\Lambda_{\widehat{\mathbf{h}}_{1}}
=-I&\ln \! \Big(\frac{\pi\sigma^2}{N}\Big)
-\frac{N}{\sigma^2}
\Big(
\widehat{\mathbf{h}}_{1}^H \widehat{\mathbf{h}}_{1}
+2\mathcal{R}\{
\xi_2 \widehat{\mathbf{h}}_{1}^H \bm{\varsigma}_0
-\xi_2 \widehat{\mathbf{h}}_{1}^H  \widehat{\mathbf{h}}_{0}  \}
\bigg.\nonumber \\
&\bigg.+|\xi_2|^2
\big\{ \widehat{\mathbf{h}}_{0}^H \widehat{\mathbf{h}}_{0}
-2\mathcal{R}\{ \bm{\varsigma}_0^H \widehat{\mathbf{h}}_{0} \}
+\bm{\varsigma}_0^H \bm{\varsigma}_0 \big\} \Big).  
\end{align}

Again, since $ \xi_2 $ is complex, it is necessary to define the real 2D vector  
$ \bm{\xi}_2=
\left[\xi_{2,\textrm{R}}, 
\xi_{2,\textrm{I}} \right]^T $ 
containing its real part and imaginary part, 
and next construct the FIM as 
$ \big[ \bm{\mathcal{F}}_{\widehat{\mathbf{h}}_{1}} (\bm{\xi}_2) \big]_{p,q}=
-\mathbb{E} \Big[
\frac{\partial^{2}\Lambda_{\widehat{\mathbf{h}}_{1}}}
{\partial \left[\bm{\xi}_{2}\right]_p  \left[\bm{\xi}_{2}\right]_q } \Big] $
with $ 1\leq p,q \leq2 $. 
Similar to $ \beta_1 $, we assume that 
$ |\beta_2|=1 $ in the sequel\footnote{Without loss of generality, it is straightforward to extend the  derivation of CRLBs into the general case with 
$ |\beta_1|,|\beta_2|\neq1 $.},   
and the aforementioned FIM is succinctly given by  
\begin{equation}
\bm{\mathcal{F}}_{\widehat{\mathbf{h}}_{1}}(\bm{\xi}_2)
=  
\left[\begin{matrix}
\frac{2NIM}{\sigma^2} 
& 0  \\
0
& \frac{2NIM}{\sigma^2}  
\end{matrix}\right].
\end{equation}
Accordingly, the CRLB on the variance of  
$ \hat{\xi}_2 $ holds   
\begin{equation}
\mathbb{E}\big[\big| \xi_2 
-\hat{\xi}_2 \big|^2 \big]
\geq \operatorname{tr} \!
\big\{( \bm{\mathcal{F}}_{\widehat{\mathbf{h}}_{1}}^{-1}
(\bm{\xi}_2) \big\}
=\frac{\sigma^2}{NIM}.
\end{equation}

To determine the bound on 
$ \hat{f}_{\textrm{d}2} $, 
the corresponding FIM (scalar) is expressed as  
\begin{equation}
\mathcal{F}_{\widehat{\mathbf{h}}_{1}}(f_{\textrm{d}2}) 
=-\mathbb{E}\!\left[
\frac{\partial^{2}\Lambda_{\widehat{\mathbf{h}}_{1}}}
{\partial f_{\textrm{d}2}^2 } \right]
=\frac{8\pi^2 N I^3 M T^2}{\sigma^2},  
\end{equation}
and thus the CRLB for $ \hat{f}_{\textrm{d}2} $  follows as 
\begin{equation}
\mathbb{E}\big[\big| f_{\textrm{d}2}
-\hat{f}_{\textrm{d}2} \big|^2 \big]
\geq \mathcal{F}^{-1}_{\widehat{\mathbf{h}}_{1}}(f_{\textrm{d}2})
=\frac{\sigma^2}{8 \pi^2 N I^3 M T^2}.
\end{equation}

\subsection{CRLBs of Path Gains and Phase Differences}
\subsubsection{Direct Link}
With the aid of (\ref{g0}), the log-likelihood function of $ \widehat{\mathbf{g}}_0 $ conditioned on $ \{\beta_1,\mathbf{d}_1\} $ is written as  
\begin{equation}
\Lambda_{\widehat{\mathbf{g}}_{0}}
=-I\ln \! \Big(\frac{\pi\sigma^2}{N}\Big)  
-\frac{N}{\sigma^2}  
\Big(
\widehat{\mathbf{g}}_{0}^H \widehat{\mathbf{g}}_{0} 
-2\mathcal{R}\{
\beta_1 \widehat{\mathbf{g}}_{0}^H \mathbf{d}_1 \}
+|\beta_1|^2 I
\Big).
\end{equation}

Once again, let us separate the real part and imaginary part of $ \beta_1 $, say 
$ \beta_{1,\textrm{R}} $ and $ \beta_{1,\textrm{I}} $, respectively,  
and define the 2D real vector  
 $ \bm{\beta}_1=
\left[\beta_{1,\textrm{R}}, 
\beta_{1,\textrm{I}} \right]^T $.  
Accordingly, consider the FIM as  
$ \big[ \bm{\mathcal{F}}_{\widehat{\mathbf{g}}_{0}} (\bm{\beta}_1) \big]_{p,q}=
-\mathbb{E} \Big[
\frac{\partial^{2}\Lambda_{\widehat{\mathbf{g}}_{0}}}
{\partial \left[\bm{\beta}_{1}\right]_p  \left[\bm{\beta}_{1}\right]_q } \Big] $
with $ 1\leq p,q \leq2 $.  
It is straightforward to realize that the FIM above also has a diagonal structure, with its diagonal entries equally given by  
\begin{equation}
\big[ \bm{\mathcal{F}}_{\widehat{\mathbf{g}}_{0}} (\bm{\beta}_1) \big]_{1,1}=
\big[ \bm{\mathcal{F}}_{\widehat{\mathbf{g}}_{0}} (\bm{\beta}_1) \big]_{2,2}=
\frac{2NI}{\sigma^2},
\end{equation}
and consequently the CRLB for $ \hat{\beta}_1 $ is obtained as 
\begin{equation}
\mathbb{E}\big[\big| \beta_1 
-\hat{\beta}_1 \big|^2 \big]
\geq \operatorname{tr}\!
\big\{( \bm{\mathcal{F}}_{\widehat{\mathbf{g}}_{0}}^{-1}
(\bm{\beta}_1) \big\}
=\frac{\sigma^2}{NI}.
\end{equation}

\subsubsection{Cascaded Link}
With given 
$ \{\beta_2,\mathbf{\Gamma},\mathbf{a}\} $ in (\ref{h}), 
the logarithm likelihood function of $ \widehat{\mathbf{h}} $ can be expressed as 
\begin{align}
\Lambda_{\widehat{\mathbf{h}}}
=-2I\ln \! \Big(\frac{\pi\sigma^2}{N}&\Big.\Big.\Big)
-\frac{N}{\sigma^2}
\Big(
\widehat{\mathbf{h}}^H \widehat{\mathbf{h}}
\Big. \nonumber \\
 \Big.
&-2\mathcal{R}\{
\beta_2 \widehat{\mathbf{h}}^H 
\mathbf{\Gamma} \widetilde{\mathbf{\Phi}} \mathbf{a} \}
+2|\beta_2|^2 IM \Big) .
\end{align}

For the case at hand, complex $ \beta_2 $ and real  
$ \{\varphi_y,\varphi_z\} $ are to be estimated.   
Let $ \beta_{2,\textrm{R}} $ and 
$ \beta_{2,\textrm{I}} $ 
denote the real part and imaginary part of $ \beta_2 $, respectively,  
and then consider the unknown four-dimension real vector
$ \bar{\mathbf{z}}=
\left[ \beta_{2,\textrm{R}}, \beta_{2,\textrm{I}},
 \varphi_y, \varphi_z \right]^T $.   
This way, the corresponding FIM is defines as
$ \big[ \bm{\mathcal{F}}_{\widehat{\mathbf{h}}} (\bar{\mathbf{z}}) \big]_{p,q}=
-\mathbb{E} \Big[
\frac{\partial^{2}\Lambda_{\widehat{\mathbf{h}}}}
{\partial \left[\bar{\mathbf{z}}\right]_p  \left[\bar{\mathbf{z}}\right]_q } \Big] $
with $ 1\leq p,q \leq 4 $,   
and take its $ (1,3) $th and $ (3,4) $th elements  as examples, which are individually expressed as  
\begin{align}
\big[ \bm{\mathcal{F}}_{\widehat{\mathbf{h}}} (\bar{\mathbf{z}}) \big]_{1,3}&=
-\frac{2N}{\sigma^2}
\mathcal{R} \!  \left\{ 
j \mathbb{E}  \big[ \widehat{\mathbf{h}}^H \big]
\mathbf{\Gamma} \widetilde{\mathbf{\Phi}}
\operatorname{diag}\{\bm{\varpi}_y\}\mathbf{a}
 \right\} \nonumber \\
&=-\frac{4NI}{\sigma^2} \beta_{2,\textrm{I}} \left\| \bm{\varpi}_y \right\|_1,
\end{align}
and that 
\begin{align}
\big[ \bm{\mathcal{F}}_{\widehat{\mathbf{h}}} (\bar{\mathbf{z}}) \big]_{3,4}&=
\frac{2N}{\sigma^2}
\mathcal{R} \!  \left\{ 
\beta_2 \mathbb{E}\big[ \widehat{\mathbf{h}}^H \big] 
\mathbf{\Gamma} \widetilde{\mathbf{\Phi}}
\operatorname{diag}\{\bm{\varpi}_y\}
\operatorname{diag}\{\bm{\varpi}_z\} \mathbf{a}
 \right\} \nonumber \\
&=\frac{4NI}{\sigma^2} \bm{\varpi}_y^T \bm{\varpi}_z.
\end{align}
Moreover, we can similarly obtain the other entries, and thus the FIM is given by (\ref{FIMl}), as shown at the bottom of last page. We see  that the FIM is a real Hermitian matrix with a non-diagonal structure, 
and it depends on $ \beta $ and the last two columns of the coefficient matrix 
$ \mathbf{\Omega} $. 
As a result, we finally determine the bounds as
\begin{equation}
\setcounter{equation}{51}
\mathbb{E} \big[
\big(\bar{\mathbf{z}}-\hat{\bar{\mathbf{z}}}\big)
\big(\bar{\mathbf{z}}-\hat{\bar{\mathbf{z}}}\big)^T
\big] \ge 
\bm{\mathcal{F}}_{\widehat{\mathbf{h}}}^{-1} (\bar{\mathbf{z}}),   
\end{equation}
and $ \hat{\beta}_2 $ is bounded by the sum of the CRLBs for its real and imaginary parts. 


\begin{figure}[t]
\centering
\includegraphics[width=1\linewidth]{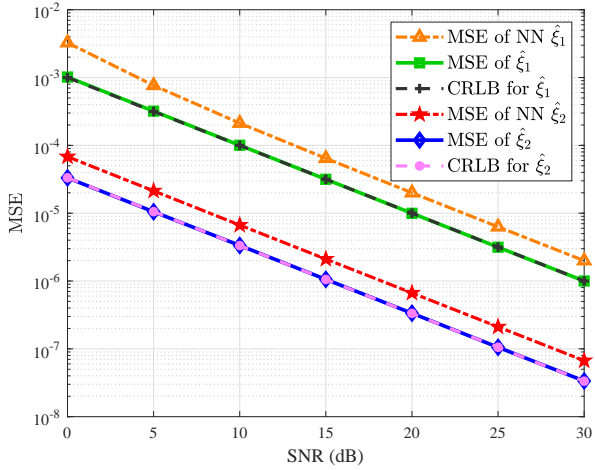}
\caption{MSEs and CRLBs versus SNR for 
$ \hat{\xi}_1 $ and $ \hat{\xi}_2 $.}
\label{fig:xi}
\end{figure}

\section{Numerical Results}\label{simulation}
This  section provides  numerical results to assess the performances of  the proposed channel  estimators for the ITS-assisted HSR network. 
Specifically, we exploit the mean squared error (MSE) as the performance metric, 
which is approximately obtained from the statistical average over more than $ 10^5 $ simulations. 
The system parameter settings of our numerical experiments  are listed as follows. 
The transmit signal-to-noise ratio (SNR) in dB is defined as $ 10\log_{10}\frac{1}{\sigma^2} $ as the result of the signal power $ P_{\textrm{s}}=1 $, and we range it from 0 to 30 dB.  
The velocity of train $ v $ is 360 km/h, carrier wavelength is assumed as $ \lambda=0.1 $ m  (corresponding to a 3 GHz carrier frequency), and the Doppler frequency shifts for the direct and cascaded links  are individually set as 
$ f_{\textrm{d}1}=901 $ Hz, 
$ f_{\textrm{d}2}=900 $ Hz. 
The quantities of subblocks in each block and pilots within every subblock are fixed to $ I=40, N=25 $, 
and the duration of every subblock is 0.01 ms. 
The path gains are 
$ \beta_1=e^{j\frac{\pi}{4}}, \beta_2=e^{j\frac{\pi}{5}}$ 
on account of the assumption 
$ |\beta_1|,|\beta_2|=1 $ in Section \ref{CRLB}, 
and the equivalent phase differences of ITS on the $ y $ and $ z $ are 
$ \varphi_y=0.08\pi, \varphi_z=0.06\pi  $. 
Finally, the number of  ITS elements is set as $ M=30 $ with $ M_y=5, M_z=6 $. 

Fig. \ref{fig:xi} illustrates the MSEs and CRLBs versus SNR for the estimators of 
$ \xi_1 $ and $ \xi_2 $, respectively, 
wherein we also depict the performances of the corresponding non-normalized (NN) LS estimators for comparison. 
We  observe that all of the MSEs and CRLBs trend downwards with the increasing  SNR. Significantly, our estimators attain their CRLBs, which demonstrates their superiority, while the constant gaps are found as for the NN estimators.  An unbiased estimator which achieves CRLB lower bound is said to be efficient. That estimator  achieves the lowest possible MSE  among all unbiased methods.  Moreover, at the same SNR, the estimation performance of $ \hat{\xi}_2 $ outperforms 
$ \hat{\xi}_1 $. 
This is because that in contrast to the direct channel, the cascaded channel is superimposed over $ M $ propagation channels through the ITS.

\begin{figure}[t]
\centering
\includegraphics[width=1\linewidth]{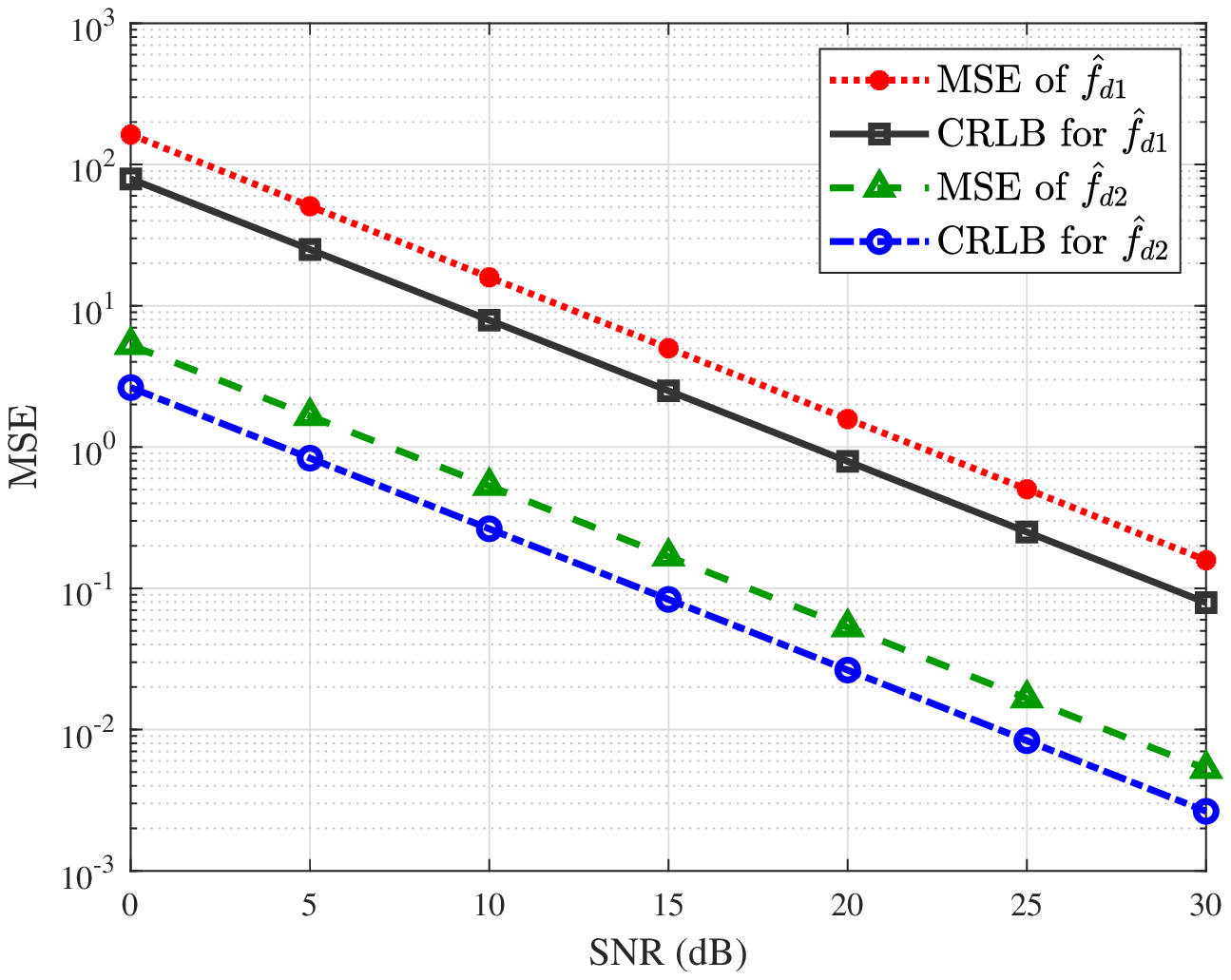}
\caption{MSEs and CRLBs versus SNR for 
$ \hat{f}_{\textrm{d}1} $ and 
$ \hat{f}_{\textrm{d}2} $.}
\label{fig:fd}
\end{figure}

Fig. \ref{fig:fd} plots the MSEs and CRLBs versus SNR for  
$ \hat{f}_{\textrm{d}1} $ and 
$ \hat{f}_{\textrm{d}2} $.   
We can readily observe that both of the Doppler shift estimators have constant performance gaps with respect to the derived CRLBs.   
We further observe that this finding is in accordance with that of the NN estimators (Fig. \ref{fig:xi}). 
This is because that our estimators of 
$ \xi_1 $ and $ \xi_2 $ 
only enhance the estimation performances of their  amplitudes based on the NN estimators; 
however, the Doppler-shift estimates are obtained from the corresponding arguments. 

\begin{figure}[t]
\centering
\includegraphics[width=1\linewidth]{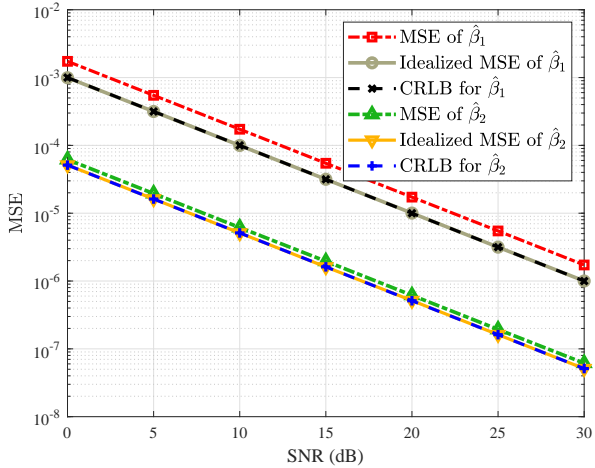}
\caption{MSEs and CRLBs versus SNR for 
$ \hat{\beta}_{1} $ and $ \hat{\beta}_{2} $.}
\label{fig:beta}
\end{figure}

Fig. \ref{fig:beta} shows the MSEs at different SNRs of 
$ \hat{\beta}_1 $ and $ \hat{\beta}_2 $ 
benchmarked by their CRLBs. 
To isolate the effects of Doppler shifts, we also estimate the path gains with perfectly known 
$ f_{\textrm{d}1} $ and $ f_{\textrm{d}2} $ as a  comparison, 
and those  estimation performances are labeled  as idealized MSEs in Fig. \ref{fig:beta}.   
We find  that the constant gaps exist between the MSEs and their corresponding CRLBs. This is because we adopt the estimates of Doppler shifts to extract  the path gains, the estimation performances suffer from the error propagation. Nonetheless, the superimposed structure of the cascaded channel contributes to narrowing the gap, since we can obtain more accurate 
$ \hat{f}_{\textrm{d}2} $ than  
$ \hat{f}_{\textrm{d}1} $.   

\begin{figure}[t]
\centering
\includegraphics[width=1\linewidth]{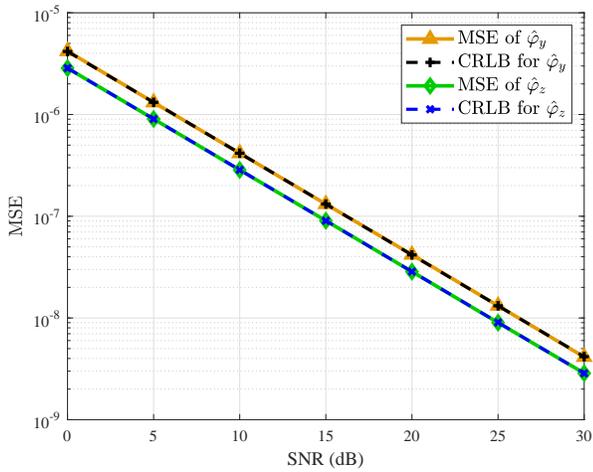}
\caption{MSEs and CRLBs versus SNR for 
$ \hat{\varphi}_{y} $ and $ \hat{\varphi}_{z} $.}
\label{fig:varphi}
\end{figure}

Fig. \ref{fig:varphi} illustrates the MSE performances versus SNR for  
$ \hat{\varphi}_y $ and $ \hat{\varphi}_z $ 
compared with the derived CRLBs. 
Unlike  Fig. \ref{fig:beta}, 
we observe that the proposed estimators of phase differences reach the CRLBs even with the existence of error spread.  
Moreover, 
 the estimator of $ \varphi_z $ achieves better performance than that of $ \varphi_y $ with respect to the same SNR. This is due to the non-square structure employed at the  ITS with $ M=5\times6 $.

\section{Conclusion}\label{conclusion}

This paper considered ITS-assisted HSR  wireless communication systems and studied their channel estimation problem. 
The channels were first modeled with physical parameters, and the two-phase transmission scheme was proposed where two pilot blocks were transmitted within each frame. 
With the refraction coefficients of the ITS as prior knowledge, the serial channel estimation algorithm was developed based on LS estimation principles. We derived the CRLB for each parameter to evaluate the performance of the corresponding estimator. Specifically, we first obtained the channel estimates, from which the channel parameters were further recovered by leveraging the relation between channels for the two pilot blocks. Numerical results demonstrated the superiority of our algorithm and corroborated the analytical derivations.

\end{document}